\documentclass[aps,preprint, prl, showpacs,nobibnotes]{revtex4-1}
\usepackage{graphicx}
\usepackage{bm,amsmath}
\usepackage{dcolumn} 
\usepackage{natbib,xcolor}

\begin{document}
\title{Extreme Magnetoresistance in Magnetic \\ Rare Earth Monopnictides}

\author{Linda Ye}
\affiliation{Department of Physics, Massachusetts Institute of Technology, Cambridge, Massachusetts 02139, USA}
\author{Takehito Suzuki}
\affiliation{Department of Physics, Massachusetts Institute of Technology, Cambridge, Massachusetts 02139, USA}
\author{Christina R. Wicker}
\affiliation{Department of Physics, Massachusetts Institute of Technology, Cambridge, Massachusetts 02139, USA}
\author{Joseph G. Checkelsky}
\affiliation{Department of Physics, Massachusetts Institute of Technology, Cambridge, Massachusetts 02139, USA}
\date{\today}
\maketitle

\textbf{The acute sensitivity of the electrical resistance of certain systems to magnetic fields known as extreme magnetoresistance (XMR) has recently been explored in a new materials context with topological semimetals.  Exemplified by WTe$_{2}$ and rare earth monopnictide La(Sb,Bi), these systems tend to be non-magnetic, nearly compensated semimetals and represent a platform for large magnetoresistance driven by intrinsic electronic structure.  Here we explore electronic transport in magnetic members of the latter family of semimetals and find that XMR is strongly modulated by magnetic order.  In particular, CeSb exhibits XMR in excess of $1.6 \times 10^{6}$\% at fields of 9 T  while the magnetoresistance itself is non-monotonic across the various magnetic phases and shows a transition from negative magnetoresistance to XMR with field above magnetic ordering temperature $T_{N}$.  The magnitude of the XMR is larger than in other rare earth monopnictides including the non-magnetic members and follows an non-saturating power law to fields above 30 T.  We show that the overall response can be understood as the modulation of conductivity by the Ce orbital state and for intermediate temperatures can be characterized by an effective medium model. Comparison to the orbitally quenched compound GdBi supports the correlation of XMR with the onset of magnetic ordering and compensation and highlights the unique combination of orbital inversion and type-I magnetic ordering in CeSb in determining its large response.  These findings suggest a paradigm for magneto-orbital control of XMR and are relevant to the understanding of rare earth-based correlated topological materials.
}

The rare earth monopnictides $RX$ crystallize in the NaCl structure (Fig. 1(a)) and exhibit a rich variety of magnetic ground states \cite{RareEarthHandBook}. In terms of electronic structure, most $RX$ compounds are known as compensated semimetals with the conduction band deriving from rare earth 5$d$ $t_{2g}$ states and valence band from pnictogen $^{3/2}p$ states, located at $\Gamma$ and $X$ points in the Brillouin zone, respectively (Fig. 1(b)) \cite{CeSbARPES}. Potential topological aspects of the electronic structure have recently been discussed, including Dirac semimetal nodes or topological insulating gaps along $\Gamma - X$ depending on the pnictogen \cite{LaSbTI, LaSbCava} (highlighted in blue) and an unusual four-fold degenerate Dirac surface state at $\bar{M}$ \cite{CeSbHasan} (projected on to $X$ shown in green).  Combining this with the $f-$electron degree of freedom suggests $RX$ may therefore be host to topological phases of correlated electrons.  This is further enriched by the reports of extreme magnetoresistance (XMR) in LaSb \cite{KasuyaTransport, LaSbCava} and LaBi \cite{KasuyaTransport, LaSbCava2, LaBi}.  While a rarity, XMR is of fundamental interest in terms of its microscopic origin and has technological relevance in magnetic sensing and related technologies \cite{XMR1, XMR2}.  The impact of strongly correlated behavior on this remarkable transport response in semimetals has thus far remained unexplored.  

The magnetism of the $RX$ compounds is distinct from that in simpler magnetic metals such as Fe, Gd, Tb and dilute magnetic semiconductors such as Mn$_x$Ga$_{1-x}$As \cite{DMS} owing to the combination of strongly localized $f$ electrons and low density, high mobility carriers from the $p$ and $d$ bands.  The NaCl structure enforces a significant interaction between the two with the principle pnictogen wave function transfer being mediated through the rare earth wave function (and vice versa) leading to a relatively wide variety of magnetic phases.  This behavior is particularly distinct for the choice of a single $f$ electron for $R=$ Ce; compounds CeP \cite{CeP}, CeAs \cite{CeAs}, and CeSb \cite{CeSbneutron} each have rich phase diagrams characterized by the mixed $f$- orbital occupation of Ce in the lattice.  We focus on strong spin-orbit CeSb here and find this magneto-orbital character has a significant impact on XMR, leading to both negative ($>$70\%) and positive ($>$1,670,000\%) magnetoresistance.  This behavior, distinct from giant magnetoresistance (GMR) in magnetic multilayers \cite{GMR} and colossal magnetoresistance (CMR) in magnetic perovskites \cite{CMR}, intertwines the semimetallic structure with magnetism and represents a potential form of magneto-orbital control of XMR.

\begin{figure}
\includegraphics[width =  \columnwidth]{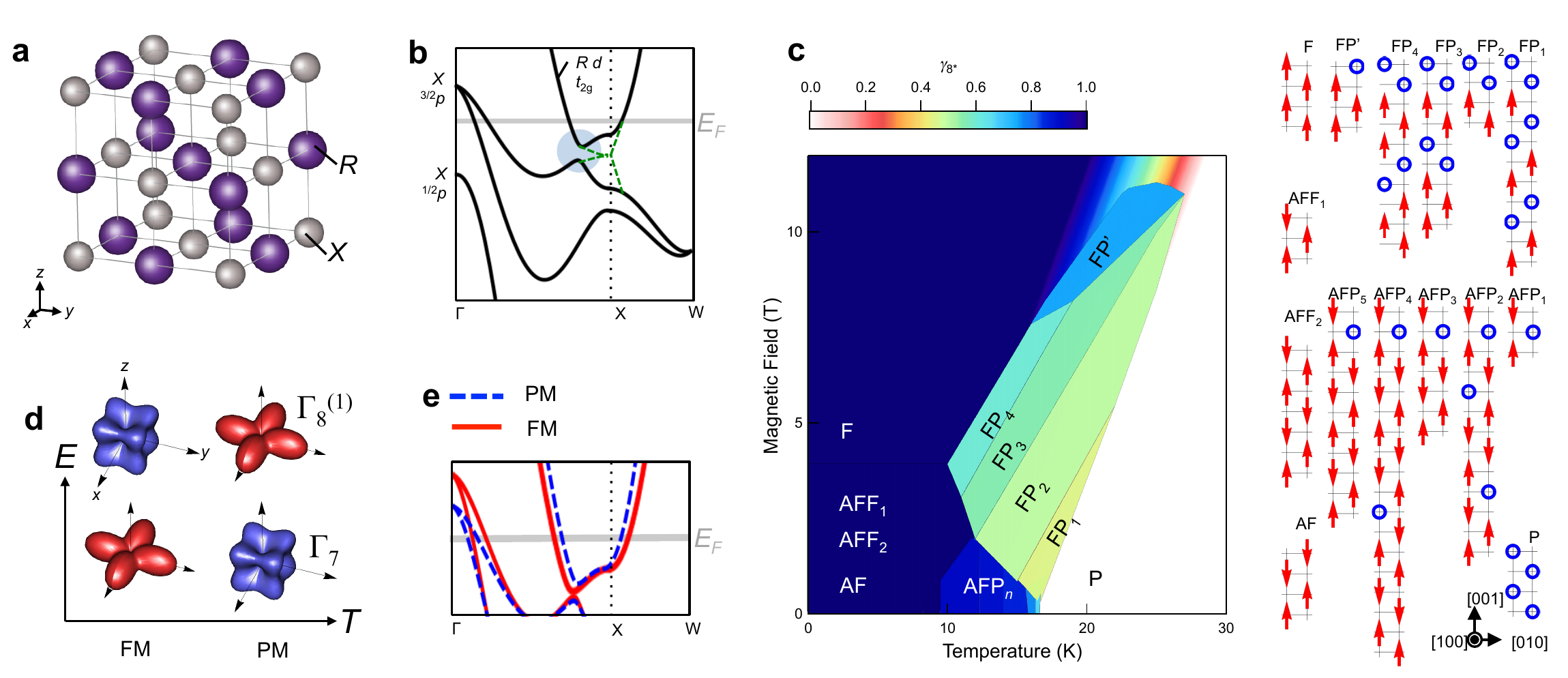}
\caption{\label{fig-1} Ye \emph{et al.}}
\end{figure}

CeSb is an unusual magnetic system, exhibiting at least 14 magnetic phases in close proximity in its magnetic field $B$ and temperature $T$ phase diagram (see Fig. 1(c)) \cite{CeSbMH}.  The primary driving force for this complexity is the interplay between the semimetallic band electrons and Ce$^{3+}$ $^1f$ states, the latter being situated near to the Fermi level $E_F$ \cite{Kasuya1}. In the high temperature paramagnetic phase, the preferred orbital state for the Ce ion is $\Gamma_{7}$ as expected from the cubic coordination (shown schematically in Fig. \ref{fig-1}(d)).  However, the $\Gamma_{8}$ states are at an energy only 3 meV higher, and in the magnetically ordered state the cruciform $\Gamma_8^{(1)}$ orbital becomes energetically favored accompanied with a shift in the electronic structure as depicted in Fig. \ref{fig-1}(e) \cite{Kasuya1} (see supplementary information).  At intermediate $T$ and $B$ a complex magnetic phase diagram arises consisting of phases built by stacking paramagnetic $\Gamma_7$ planes and ferromagnetic $\Gamma_8^{(1)}$-like planes as shown in Fig. 1(c).  Neutron \cite{Boucherle} and X-ray \cite{Xray} scattering measurements have mapped the orbital content of these phases (the $\Gamma_8^{(1)}$-like planes are reported to be composed of planar orbitals that are close to a $J_z=|\pm 5/2\rangle$ fully polarized state we hereafter refer to as $\Gamma_{8^{\star}}$, see supplementary materials); the color scale in Fig. \ref{fig-1}(c) reflects the $\Gamma_{8^{\star}}$ occupation $\gamma_{8^{\star}}$.  Among isostructural cerium monopnictides, the phase diagram of CeSb uniquely hosts antiferromagnetic (AF), antiferro-ferromagnetic (AFF$_{n}$), ferromagnetic (F), antiferro-paramagnetic (AFP$_{n}$), ferro-paramagnetic (FP$_{n}$), and paramagnetic (P) phases \cite{CeBi,CeP,Pressure,LaCeSb}.

\begin{figure}
\includegraphics[width =  \columnwidth]{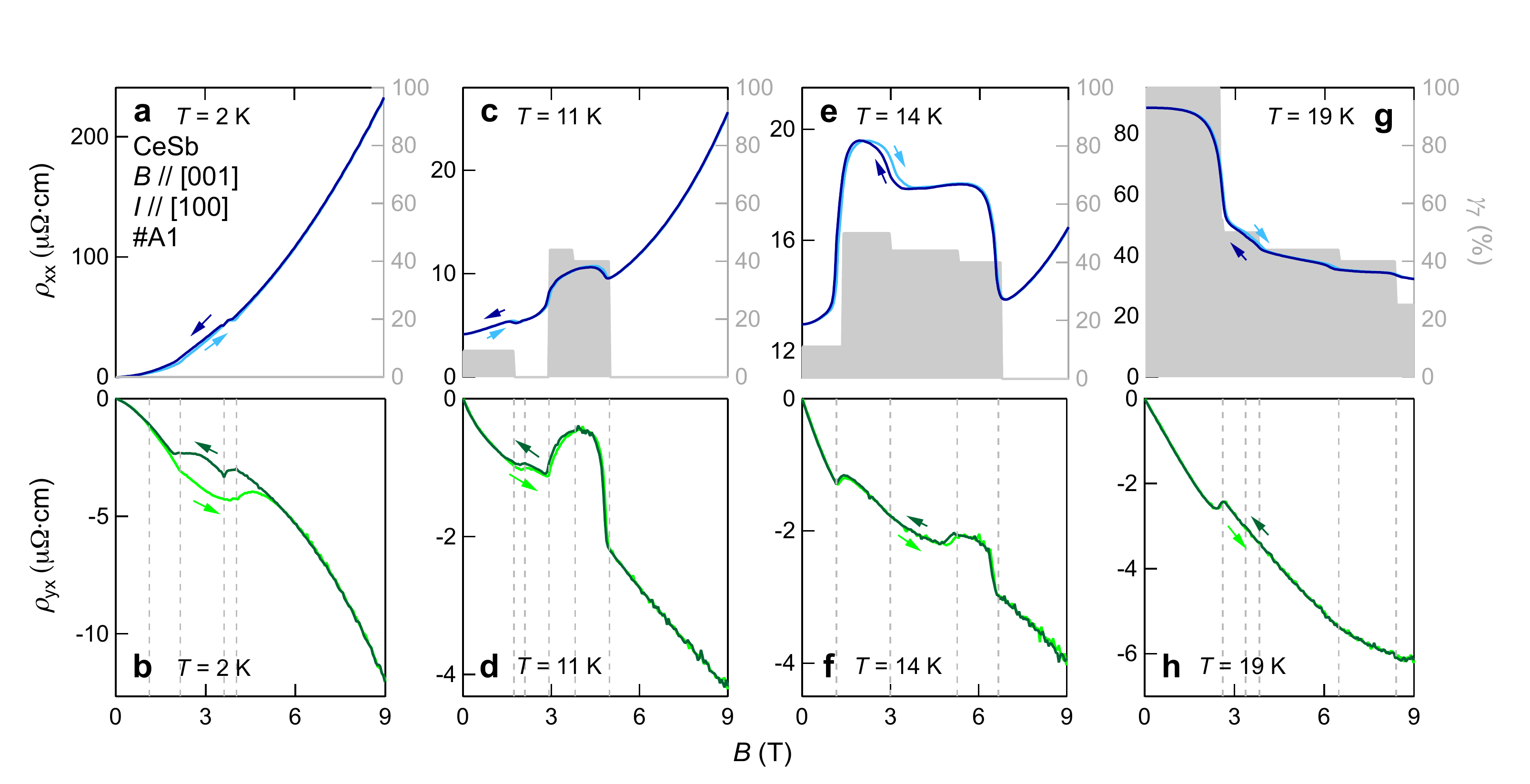}
\caption{\label{fig-2}  Ye \emph{et al.}}
\end{figure}

We first examine the longitudinal resistivity $\rho_{xx}$ and transverse resistivity $\rho_{yx}$ as a function of magnetic induction $B$ at different characteristic temperatures $T$ in the phase diagram.  We note that $B=\mu_{0}(H+(1-N)M)$ is corrected for demagnetization effects with the demagnetization factor $N$ calculated from the sample dimensions and magnetization $M$ measured separately (see supplementary materials).  Starting with sample A1 at $T=$ 2 K in Fig. \ref{fig-2}(a), $\rho_{xx}(B)$ shows a rapid positive magnetoresistance reaching $233,500$\% of its zero field value at $B=$ 9 T.  This XMR behavior is comparable to that seen in WTe$_{2}$ \cite{XMR1}, LaBi \cite{LaSbCava2, LaBi}, and LaSb \cite{LaSbCava}, and significantly larger than that reported previously in CeSb \cite{R1,R2,R3,CanfieldCeSb} ($< 10,000\%$) and other magnetic $RX$ (see supplementary section S4).  Kinks in $\rho_{xx}(B)$ are noticeable at intermediate $B$ corresponding to the magnetic phase boundaries between AF, AFF$_{n}$, and F states in Fig. \ref{fig-1}(c).  This is also seen in $\rho_{yx}(B)$ shown in Fig. \ref{fig-2}(b), where the vertical lines denote the phase boundaries observed on decreasing $B$.  A significant hysteresis in both $\rho_{xx}(B)$ and $\rho_{yx}(B)$ is observed.  The hysteresis is found to be sample dependent, similar to that reported in previous magnetization studies \cite{CeSbMH}.   

While at $T=2$ K four different phases are stable at different $B$, for all $B$ these phases are fully composed of $\Gamma_{8^{\star}}$ and thus have pure $\Gamma_{7}$ layer volume fraction $\gamma_{7} = 0\%$. Upon increasing to $T=11$ K, the AFP$_{n}$, FP$_{n}$ with mixed orbital character enter the phase diagram.  Here, as shown in Fig. \ref{fig-2}(c) the XMR response is weakened (538.5\% at $B=9$ T) and clear non-monotonic behavior in $\rho_{xx} (B)$ is observed, with regions of both positive and negative $d\rho_{xx}/dB$.  Also plotted in gray is $\gamma_{7} = 100\% - \gamma_{8^{\star}}$; a correlation between intermediate regions of enhanced $\rho_{xx}$ and $\gamma_7$ is apparent.  We expand on this below.  The Hall response (Fig. \ref{fig-2}(d)) is also sensitive to the magnetic phase boundaries, with a significant drop in magnitude in the FP$_{n}$ phases where $\gamma_7 \neq 0$.  

At higher $T = 14$ K the same $\Gamma_{7}$ rich phases occupy a wider range of field and significant positive magnetoresistance is observed only for $B > 6$ T in the F state (Fig. \ref{fig-2}(e)) with discontinuities in $\rho_{yx}(B)$ appearing at the phase boundaries (Fig. \ref{fig-2}(f)).  At $T=19$ K, above the zero field magnetic ordering temperature $T_{N} = 16$ K, positive magnetoresistance is absent up to 9 T (Fig. \ref{fig-2}(g)), while at sufficient $B$ the system transitions from the P phase to FP$_{n}$ with features in $\rho_{yx}(B)$ apparent at the critical values of $B$ (Fig. \ref{fig-2}(f)). The linear Hall effect in the P phase corresponds to a single band carrier number of $5.0 \times10^{20}$/cm$^3$ or 0.046 e$^{-}$/Ce. We note that the non-linear $\rho_{yx}(B)$ in the low temperature F phase resembles that observed at lowest $T$ in LaSb \cite{KasuyaTransport} and LaBi \cite{LaBi}; a multi-band model must be incorporated to fully account for the behavior \cite{CeSbdHvA}. More broadly, at these elevated temperatures the complex evolution of $\rho_{xx}(B)$ correlates with $\gamma_{7}$, suggestive of a connection between the orbital content and the conductivity of the system.

\begin{figure}
\includegraphics[width = \columnwidth]{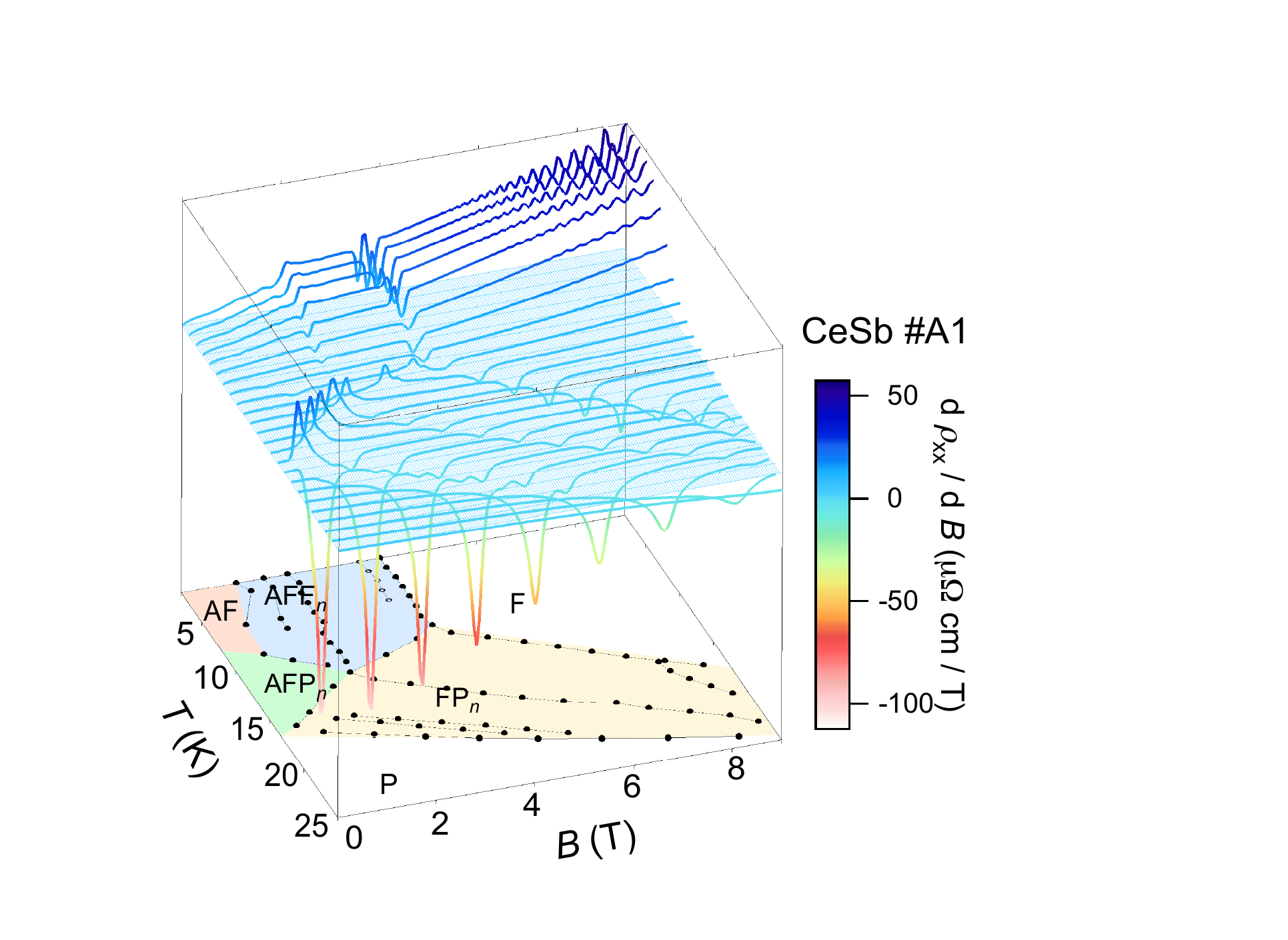}
\caption{\label{fig-3}  Ye \emph{et al.}}
\end{figure}

The evolution of the magnetoresistance is summarized in Fig. \ref{fig-3} where $d\rho_{xx} / dB$ is plotted.  Here the large, non-saturating magnetoresistance corresponding to XMR can be seen at low $T$ with a superimposed Shubnikov-de Haas oscillation.  The oscillation (frequency $= 213.2 \pm 0.5$ T) corresponds to the $k_x-k_y$ cross section of the $X_Z$ electron pocket \cite{CeSbdHvA}. For higher $T > T_{N}$ there are regions of striking negative magnetoresistance, reaching magnitudes of 100 $\mu \Omega$ cm / T at the phase boundary between P and FP phases.  The sharp features are suppressed in transitions between FP$_n$ phases, and an overall negative magnetoresistance is observed, reaching a magnitude of $72\%$ at $T=17$ K.  It is noteworthy that this negative magnetoresistance differs from conventional field suppression of magnon scattering which follows a $B$-linear trend and is typically at the percent level at comparable $B$ \cite{Raquet}.

As an aside we note that using the sharp features in $d\rho_{xx} / dB$ it is possible to construct the phase diagram of CeSb purely from transport.  This is shown projected in the $B-T$ plane in Fig. \ref{fig-3}.  Closed circles are features that reproduce those seen in $M$ (see supplementary materials).  Interestingly, we see an additional feature that develops in the AFF$_{1}$ phase in decreasing $B$ not previously reported in magnetization studies (open circles).  This may correspond to a previously unidentified phase that further enriches the phase diagram of CeSb.  

\begin{figure}
\includegraphics[width = 0.6 \columnwidth]{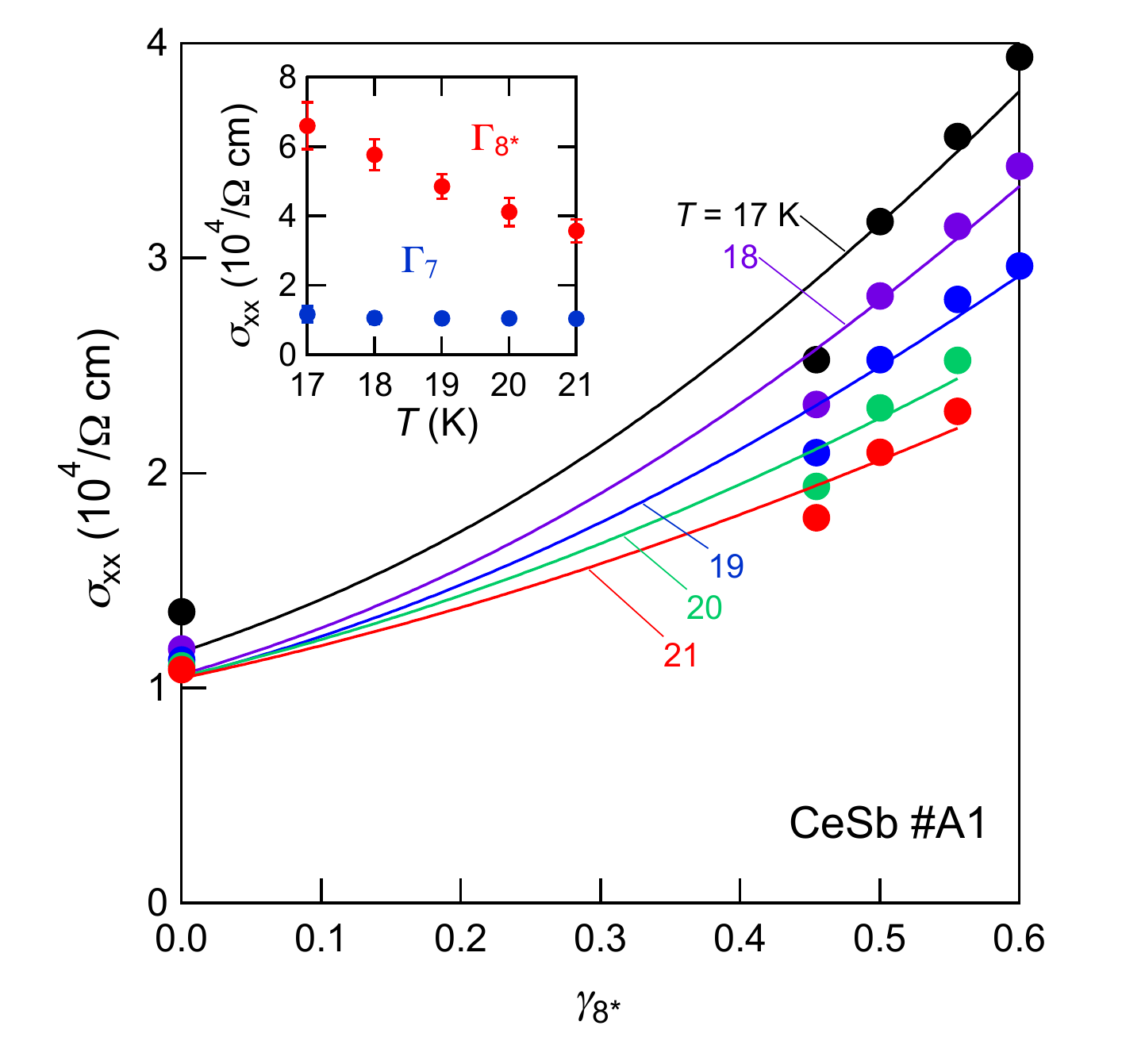}
\caption{\label{fig-4} Ye \emph{et al.}}
\end{figure}

A detailed comparison of transport to the orbital content across the phase diagram is motivated by recent X-ray analysis demonstrating the evolution of the localized $f$ wave function from $\Gamma_{8^{\star}}$ to $\Gamma_7$ with increasing $T$ across the zero field AF, AFP$_{n}$, and P states \cite{Xray}.  As discussed above, compared to paramagnetic $\Gamma_{7}$, $\Gamma_{8^{\star}}$ enhances hopping between the neighboring Sb sites in the plane (Fig \ref{fig-1}(d)) and therefore may be expected to lead to enhanced conductivity.  This is qualitatively consistent with the correlation of $\gamma_{7}$ and enhanced $\rho_{xx}$ in Figs. \ref{fig-2}(c), \ref{fig-2}(e) and \ref{fig-2}(g).  To further quantify this, we model the system as a binary mixture with relatively low and high conductivity component layers composed of the $\Gamma_{7}$ and $\Gamma_{8^{\star}}$ orbitals, respectively.  A common approach to conductivity in two component mixtures first developed for composites \cite{Landauer}, and later applied to systems ranging from superconductors \cite{EM_SC} to CMR manganites \cite{EM_CMR}, is that of the effective medium.  The underlying assumption is that a given region can be considered to be surrounded by a medium with uniform conductivity characteristic of the mixture \cite{MsLachlan}.  Denoting the  conductivities of $\Gamma_{7}$ and $\Gamma_{8^{\star}}$ regions as $\sigma_{7}$ and $\sigma_{8^{\star}}$, respectively, in this model the total conductivity can be written as
\begin{equation}
\sigma_{xx}(\gamma_{8^{\star}}) = \sigma_{c} + \sqrt{\sigma_{c}^{2}/4 + 8\sigma_{7}\sigma_{8^{\star}}}, 
\label{EM}
\end{equation}
where $\sigma_{c} = \sigma_{7}(2-3\gamma_{8^{\star}}) + \sigma_{8}(3\gamma_{8^{\star}}-1)$ (see supplementary materials).  Calculating the conductivity as $\sigma_{xx} = \rho_{xx} / (\rho_{xx}^{2} + \rho_{yx}^{2})$ we estimate $\sigma_{xx}$ at each phase as its value at the central $B$ between the phase boundaries.  For simplicity we focus on the region above $T_{N}$ where $\gamma_{7}$ evolves most rapidly.  These values are shown as a function of $\gamma_{8^{\star}}$ in Fig. \ref{fig-4} as circles with the fit to Eq. (\ref{EM}) as the solid line.  For metal-insulator mixtures this curve takes a divergent shape reflecting the percolative transition between the two end phases \cite{EM_CMR}.  Here the dependence is more gentle, suggesting components of comparable conductivity.  

The orbital-dependent conductivity values found at each $T$ in the inset of Fig. \ref{fig-4}, indicating a conductivity ratio of approximately 6 at $T=17$ K that diminishes on warming. While both orbital dependent conductivities are metallic, the ferromagnetic $\Gamma_{8^{\star}}$ contribution rises rapidly as $T$ is reduced, eventually leading to the high conductivity XMR state at low $T$.  While this confirms the above observation of enhanced resistivity in $\Gamma_{7}$ rich states, it further suggests a manner of control for XMR by the orbital degree of freedom.  If the orbital content could be manipulated at low $T$ this suggests XMR could be similarly modulated. That XMR would be absent at low $T$ if the orbital content were modified is supported by magnetotransport reports in CeP which has a $\Gamma_{7}$ rich ground state at low $T$ and shows negligible XMR at similar fields \cite{CeP_MR}.  Application of pressure may therefore be an effective manner to tune XMR as positive pressure is known to suppress $\Gamma_{7}$ in CeSb \cite{Pressure} and negative (chemical) pressure via La doping acts in the opposite fashion \cite{LaCeSb}. Alternatively, epitaxial thin films grown on appropriate substrates may realize materials with strain-controlled XMR.  

\begin{figure}
\includegraphics[width =  \columnwidth]{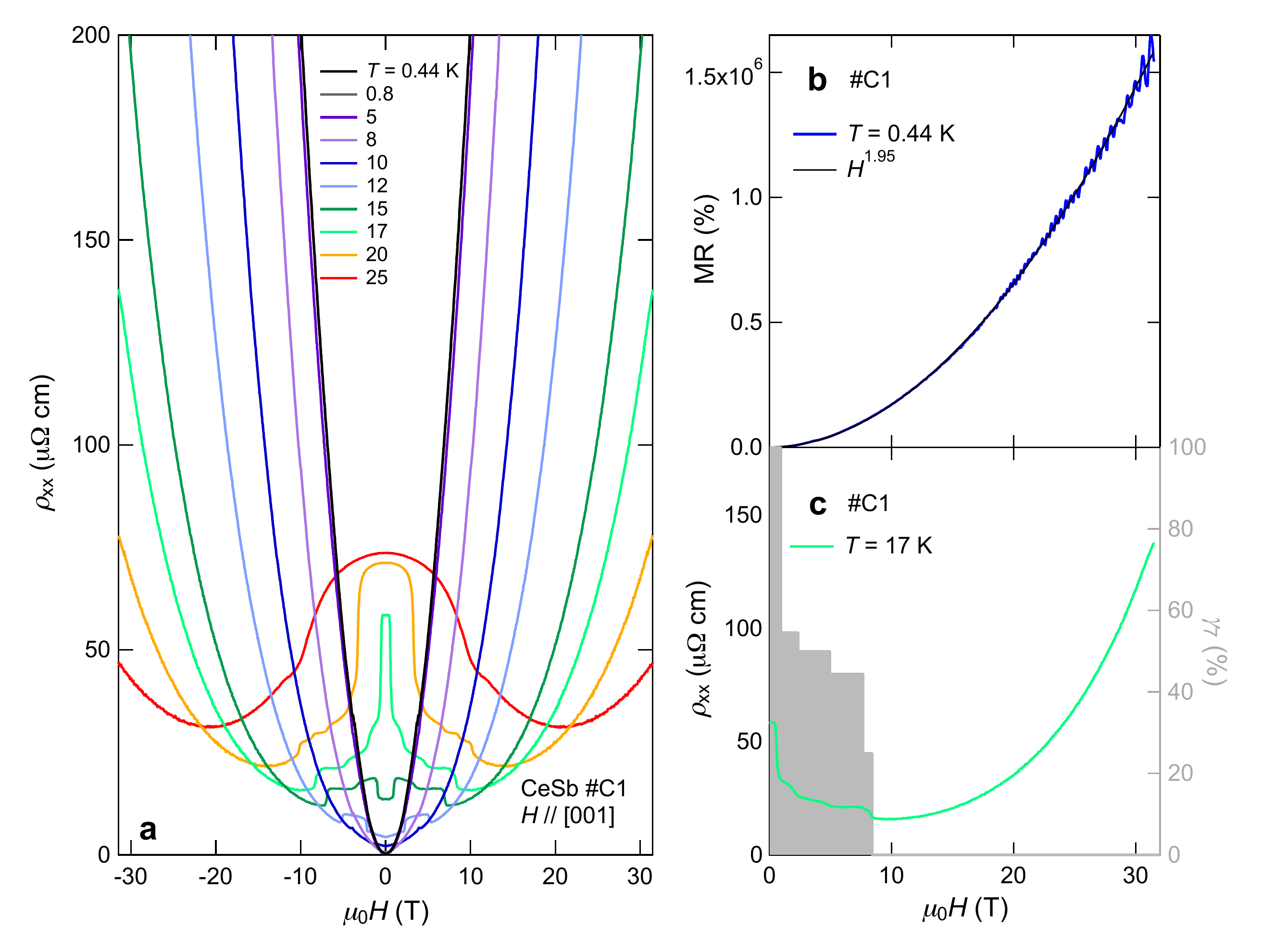}
\caption{\label{fig-5} Ye \emph{et al.}}
\end{figure}

The non-saturating nature of the XMR in CeSb is observed with application of larger magnetic fields. As shown in Fig. \ref{fig-5}(a), sample C1 measured to fields above 30 T shows a similar crossover pattern from negative to positive magnetoresistance at intermediate $T$ and sharply increasing XMR at the lowest $T$.  As shown in Fig. \ref{fig-5}(b), for $T=0.44$ K the MR is in excess of $1,500,000\%$ at the highest fields and is well described by a 1.95 power law without sign of saturation (pronounced SdH oscillations are observed, see supplementary section S5).  Large magnetic field also demonstrates the correlation of positive MR with field induced $\Gamma_{8^{\star}}$ planar orbitals \cite{Kasuya1}.  For $T=17$ K, $\gamma_{7}$ drops to zero near 9 T after which a large, non-saturating MR emerges (see Fig. \ref{fig-5}(c)).  

\begin{figure}
\includegraphics[width =  \columnwidth]{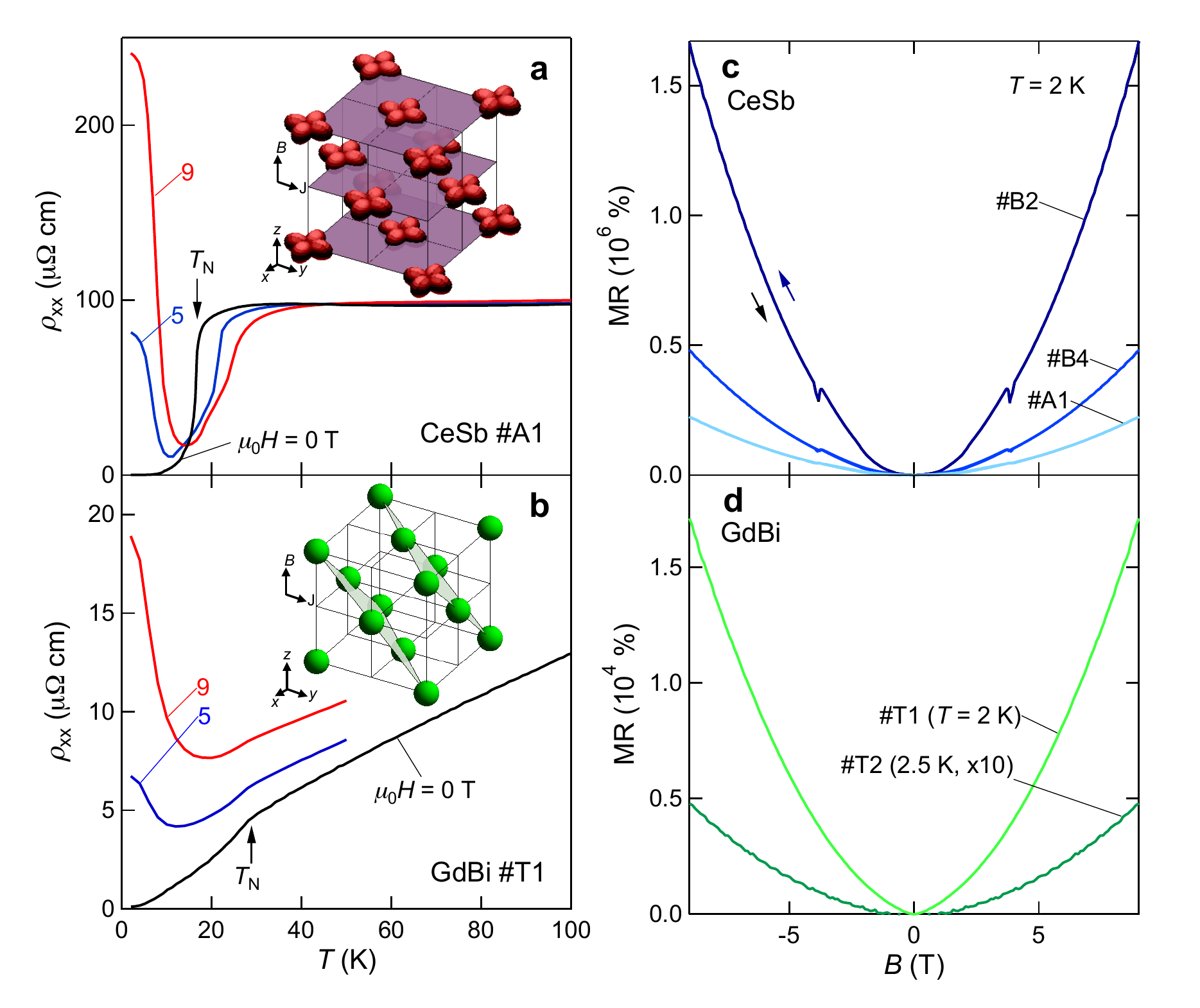}
\caption{\label{fig-6} Ye \emph{et al.}}
\end{figure}

To further elucidate the origin of XMR at low $T$ we compare the response of CeSb to $RX$ compound GdBi, which is expected to be similar in electronic structure but is orbitally quenched.  The overall metallicity and field response is shown in Fig. \ref{fig-6}(a) and Fig. \ref{fig-6}(b) for CeSb sample S1 and GdBi sample T1, respectively.  For CeSb, $\rho_{xx}$ drops dramatically below $T_{N}$ reaching a value of 1 n$\Omega$ m at $T=2$ K (Residual resistivity ratio RRR = 1017), while application of $B$ induces XMR.  For GdBi, the behavior is similar with a drop in $\rho_{xx}$ to 1.2 n$\Omega$ m at $T=2$ K (RRR = 255) and XMR approximately one order of magnitude smaller at $B = 9$ T.  XMR for CeSb samples A1, B2, and B4 is shown in Fig. \ref{fig-6}(c), the largest of which reaches $1,672,200\%$ at 9 T (RRR = 2726 and residual resistivity 77 n$\Omega$cm).  This is larger than any previous report in the $RX$ family, including the non-magnetic LaBi and LaSb.  For GdBi, XMR is observed as shown in Fig. \ref{fig-6}(d), reaching values of $17,125\%$ (previous reports of GdSb have reported similar values of $2,300\%$ \cite{GdX_MR}).  Whereas for CeSb multiband fitting is complicated by the various field induced transitions, GdBi remains in an antiferromagnetic state ($T_{N} = 28$ K) up to $B=31$ T.  In this case multiband fitting indicates a significant enhancement of mobility below $T_{N}$ and nearly compensated state at the lowest $T$ (see supplementary materials).  We therefore suggest that XMR in magnetic $RX$ systems share a common origin with non-magnetic La$X$ below $T_{N}$ where magnetic scattering suppressed.  We note it recently been discussed in the context of YSb that both exact compensation and moderate compensation with mobility mismatches may support this behavior \cite{YSb}.

The XMR in CeSb exceeds even that reported in its non-magnetic analogs, which is unexpected from the viewpoint of the additional disorder associated with the magnetic degree of freedom.  We hypothesize that this behavior is rooted in the orbitally-ordered ground state of CeSb.  In particular, the planar orbital favored in the magnetic ground state provides highly mobile carriers that populate the planes formed by the type-I ordering (see inset of \ref{fig-6}(a)).  It is noteworthy that the planar orbital is favored in CeSb despite the preference for the $\Gamma_7$ orbital shape in the cubic crystal field of the NaCl structure and that an unusually large magnetic anisotropy pins the moments normal to the ordered planes \cite{Kasuya1}. This is not the case, for example, for NdSb \cite{R3,NdSb}. Additionally, as type-II ordering is favored for $RX$ heavier than Eu$X$ \cite{R2}, we suggest that CeSb may realize a unique combination of orbital and magnetic ordering that gives rise to its large XMR in this configuration.  This can be contrasted with GdBi, which shows moderate XMR here and has spherical orbitals supporting a type-II antiferromagnetism (see inset of \ref{fig-6}(b)).  Theoretical work may allow prediction of significant XMR in other magnetic $RX$ and related compounds along these lines.  

The presence of XMR in rare earth monopnictides appears to be a ubiquitous phenomenon originating from their common semimetallic band structure.  The use of rare earth elements beyond La, Y, and Lu introduces correlation effects into these systems that modulate XMR.  The study here demonstrates how the anomalous ordering of crystal field states in CeSb allows this tuning with moderate $B$ and $T$.  While electronic structure calculations in the various magnetic ground states of CeSb are challenging, it is noteworthy that previous calculations in the F state show bands with the character of type II Weyl points in the vicinity of the Fermi level \cite{Fcalc}, indicating the possible role of topological features in these systems.  Furthermore, it can be expected that magnetic order may introduce exchange effects to produce magnetically induced Weyl points for the inverted gap $\Gamma - X$ direction as have been discussed for half-Heusler systems \cite{GdPtBi1, GdPtBi2}.  Further theoretical work is needed to confirm whether such scenarios occur and to more broadly understand the underlying electronic structure in these magnetic $RX$ systems and their potential for magneto-orbitally modified XMR.

\section{Methods}

Single crystals of CeSb are grown using a Sn-flux method \cite{Canfield,CanfieldCeSb} from Ce (Ames Laboratory \cite{Ames}, 99.99\%), Sb (Alfa Aesar, 99.999\%), Sn (Alfa Aesar, 99.995\%) powders. They are mixed with atomic ratio Ce:Sb:Sn = 1:1:20, put in alumina crucible and sealed in quartz tube back filled with 150 torr Ar gas. The raw materials are first heated to 1050$^{\circ}$C and slowly cooled to 750 $^{\circ}$C, at which point centrifuge separation of CeSb crystals from the Sn flux is performed.  Single crystals of GdBi are gown using a Bi self flux method \cite{Canfield} from Gd (Alfa Aesar, 99.9\%) and Bi (Alfa Aesar, 99.999\%).  They are mixed with atomic ratio Gd:Bi = 18.5:81.5, put in alumina crucible and sealed in quartz tube. The raw materials are first heated to 1100$^{\circ}$C and slowly cooled to 950 $^{\circ}$C followed by 4 days of annealing, at which point centrifuge separation of GdBi crystals from the Bi flux is performed.  In both cases sub-cm size rectangular crystals are obtained and oriented with single crystal diffraction.  Transport properties are measured in a commercial cryostat with a superconducting magnet. The magnetic field is applied along [001] and current flows along [100].  Field symmetrization/anti-symmetrization is performed on time-reversed field sweeps (up and down) to calculate the longitudinal/transverse resistivity and eliminate electrical pickup from contact misalignment.  Magnetization is measured with a commercial SQUID magnetometer. Transport measurements at the National High Magnetic Field Laboratory are performed in a $^3$He cryostat in Cell-9 with a four probe method.

\hfill

\textbf{Acknowledgments}
We are grateful to L. Fu and T. Kurumaji for fruitful discussions. This research is funded in part by the Gordon and Betty Moore Foundation EPiQS Initiative, Grant GBMF3848 to J.G.C., material development by NSF grant DMR-1554891, and instrumentation development with ARO grant W911NF-16-1-0034. L.Y. acknowledges support by the STC Center for Integrated Quantum Materials, NSF Grant No. DMR-1231319 and by the Tsinghua Education Foundation. J.G.C. acknowledges support from the Bose Fellows Program at MIT. A portion of this work was performed at the National High Magnetic Field Laboratory, which is supported by National Science Foundation Cooperative Agreement No. DMR-1157490, the State of Florida, and the US Department of Energy. 

\textbf{Author Contributions}
L. Y., T. S., and C. R. W. grew the single crystals. L. Y. and T. S. characterized the materials, performed transport and magnetic measurements, and analyzed the data.  All authors contributed to writing the manuscript.  J. G. C. supervised the project.  

\newpage

\textbf{Figure 1. Crystalline and Electronic Structure of $RX$ and CeSb.}  \textbf{a} Rocksalt (NaCl) structure of rare earth monopnictides $RX$.  \textbf{b} Schematic electronic structure for $RX$ with bands due to the pnictogen and rare earth labeled.  Topologically inverted bands (highlighted in blue) and surface bands (green) have recently been discussed.  \textbf{c} Phase diagram for CeSb.  The magnetic structure for the Ce layers is shown to the right and molar population of magnetic $\Gamma_{8^{\star}}$ orbital $\gamma_{8^{\star}}$  shown in the color scale.  \textbf{d} Relative energies and orbital structure of the  crystal field doublets $\Gamma_7$ and $\Gamma_8^{(1)}$ in the (001) in-plane ferromagnetic (FM) and paramagnetic (PM) states. \textbf{e} Schematic electronic structure for CeSb in the FM and PM states.  \\

\textbf{Figure 2. Electronic Transport in CeSb.} \textbf{a} Longitudinal resistivity $\rho_{xx}(B)$ for CeSb at $T=2$ K.  Measurements with field sweeps in both directions are shown (sweep direction labeled by arrows).  The right hand axis labels the $\Gamma_{7}$ orbital occupation $\gamma_7$ expected from the magnetic structure.  \textbf{b} Transverse resistivity $\rho_{yx}(B)$ at $T=2$ K. Sweeps up and down are both shown (sweep direction labeled by arrows). The vertical dashed lines mark the magnetic transitions expected on decreasing magnetic field.   \textbf{c-h} Transport at $T=11$, 14, and 19 K.  \\

\textbf {Figure 3. Field derivative of Resistivity.}  Development of $d\rho_{xx}/dB$ for field scans for different $T$ between 2 K and 25 K with decreasing $B$.  The sharp features are projected on the $B-T$ plane and reproduce the phase diagram obtained from magnetization (filled circles) with new features observed (open circles).  \\

\textbf{Figure 4. Effective Medium Model.}  Conductivity $\sigma_{xx}$ at different $\Gamma_{8^{\star}}$ orbital population $\gamma_{8^{\star}}$ from transport results (filled circles).  The solid line at each temperature is a fit to an effective medium model with two components (see text).  The inset shows the temperaure $T$ dependence of the two components of conductivity associated with the different orbital populations.  \\

\textbf {Figure 5. High Field Magnetotransport of CeSb.}  \textbf{a} Magnetic field dependence of longitudinal resistivity $\rho_{xx}$ for CeSb sample C1 up to 31.5 T \textbf{b} Low temperature $T=0.44$ K magnetotransport response fit to 1.95 power law.  \textbf{c} Intermediate temperature $T=17$ K magnetotransport plotted with pure $\Gamma_{7}$ layer volume fraction $\gamma_{7}$.  \\

\textbf{Figure 6. Comparison of CeSb and GdBi.} \textbf{a} Temperature dependence of longitudinal resistivity $\rho_{xx}(T)$ at different applied magnetic fields $\mu_{0}H$ for CeSb with the ordering temperature $T_{N}\approx 16$ K shown. \textbf{b} $\rho_{xx}(T)$ at different $\mu_{0}H$ for GdBi with $T_{N}\approx 28$ K labeled. The insets in a and b show the orbital shape and antiferromagnetic ordering planes with respect to the current/magnetic field directions for the magnetic ground states of CeSb and GdBi. \textbf{c} Magnetoresistance (MR) defined as $\rho_{xx}(B) / \rho_{xx}(B=0) - 1$ at $T=2$ K for CeSb crystals A1, B2, and B4. \textbf{d} Magnetoresistance for GdBi crystals T1, T2.
\clearpage

\clearpage

\clearpage

\clearpage

\clearpage

\clearpage

\begin{thebibliography}{99}
\bibitem{RareEarthHandBook} Hulliger, F. \emph{Handbook on the Physics and Chemistry of Rare Earths}, \textbf{4}, 153-236 (1979).

\bibitem{CeSbARPES} Takayama, A., Souma, S., Sato, T., Arakane, T., \& Takahashi, T.  Magnetic phase transition of CeSb studied by low-Energy angle-resolved photoemission spectroscopy. \emph{J. Phys. Soc. Jpn.} \textbf{78}, 073702 (2009).

\bibitem{LaSbTI} Zeng, M., Fang, C., Chang, G., Chen, Y. A., Hsieh, T., Bansil, A., Lin, H. \& Fu, L. Topological semimetals and topological insulators in rare earth monopnictides. arXiv:1504.03492 (2015).

\bibitem{LaSbCava} Tafti, F. F., Gibson, Q. D., Kushwaha, S. K., Haldolaarachchige, N. \& Cava, R. J. Resistivity plateau and extreme magnetoresistance in LaSb. \emph{Nat. Phys.} \textbf{12}, 272-277 (2016).

\bibitem{CeSbHasan} Alidoust, N. \emph{et al.}. A new form of (unexpected) Dirac fermions in the strongly-correlated cerium monopnictides. arXiv:1604.08571 (2016).

\bibitem{KasuyaTransport} Kasuya, T., Sera, M., Okayama, Y. \& Haga, Y.  Normal and anomalous Hall effect in CeSb and CeBi. \emph{J. Phys. Soc. Jpn.} \textbf{65}, 160-171 (1995).

\bibitem{LaSbCava2} Tafti, F. F., Gibson, Q. D., Kushwaha, S. K., Krizan, J. W., Haldolaarachchige, N. \& Cava, R. J. Temperature−field phase diagram of extreme magnetoresistance. \emph{Proc. Natl. Acad. Sci.} \textbf{113}, E3475-E3481 (2016).

\bibitem{LaBi} Sun, S., Wang,Q., Guo, P.-J., Liu, K. \& Lei, H. Large magnetoresistance in LaBi: origin of field-induced resistivity upturn and plateau in compensated semimetals. \emph{New J. Phys.} \textbf{18}, 082002 (2016).

\bibitem{XMR1} Ali, M. N. \emph{et al.} Large, non-saturating magnetoresistance in WTe$_{2}$. \emph{Nature} \textbf{514}, 205-208 (2014).

\bibitem{XMR2} Lenz, J. E. A review of magnetic sensors. \emph{Proc. IEEE} \textbf{78}, 973–989 (1990).

\bibitem{DMS} Dietl, T. A ten-year perspective on dilute magnetic semiconductors and oxides. \emph{Nat. Mater.} \textbf{9}, 965-974 (2010).

\bibitem{CeP} Terashima, T., Uji, S., Aoki, H., Perenboom, J. A. A. J., Haga, Y.,  Uesawa, A., Suzuki, T., Hill, S. \& Brooks, J. S. Successive metamagnetic transitions and magnetoresistance in the low-carrier-density strongly correlated electron system CeP. \emph{Phys. Rev. B} \textbf{58}, 309-313 (1998).

\bibitem{CeAs} Komorita, K., Kido, G., Nakagawa, Y., Kwon, Y. S. \& Suzuki, T. Magnetic phase transition and Shubnikov-de Haas effect of the Kondo system CeAs in high magnetic fields. \emph{ J. Magn. Magn. Mater.} \textbf{104-107}, 1241-1242 (1992).  

\bibitem{CeSbneutron} Rossat-Mignod, J., Burlet, P., Villain, J., Bartholin, H., Tcheng-Si, W., Florence, D. \& Vogt,O. Phase diagram and magnetic structures of CeSb. \emph{Phys. Rev. B} \textbf{16}, 440-461 (1977).

\bibitem{GMR} Baibich, M. N., Broto, J. M., Fert, A., Nguyen Van Dau, F., Petroff, F., Etienne, P., Creuzet, G., Friederich, A., \& Chazelas, J.  Giant magnetoresistance of (001)Fe/(001)Cr magnetic superlattices. \emph{Phys. Rev. Lett.} \textbf{61}, 2472-2475 (1988).

\bibitem{CMR} Jin, S., McCormack, M., Tiefel, T. H. \& Ramesh, R. Colossal magnetoresistance in La-Ca-Mn-O ferromagnetic thin films. \emph{J. Appl. Phys.} \textbf{76}, 6929-6933 (1994).

\bibitem{CeSbMH} Vogt, O. \& Mattenberger, K. The extraordinary case of CeSb. \emph{Physica B} \textbf{215}, 22-26 (1995).

\bibitem{Kasuya1} Takahashi, H. \& Kasuya, T.  Anisotropic $p$-$f$ mixing mechanism explaining anomalous magnetic properties in Ce monopnictides. IV. Ferromagnetic state. \emph{J. Phys. C} \textbf{18}, 2731-2744 (1985).

\bibitem{Boucherle} Boucherle, J. X., Delapalme, A., Howard, C.J., \& Rossat-Mignod, J.  Polarized neutron determination of the ground state of Ce$^{3+}$ in CeSb.  \emph{Physica B} \textbf{102}, 253-257 (1980).

\bibitem{Xray} Iwasa, K., Hannan, A., Kohgi, M. \& Suzuki, T. Direct observation of the modulation of the 4$f$-electron orbital state by strong $p$-$f$ mixing in CeSb. \emph{Phys. Rev. Lett.} \textbf{88}, 207201 (2002).

\bibitem{Pressure} Hannan, A., Okayama, Y., Osakabe, T., Kuwahara, K. \& Kohgi, M. Unusual electronic state of the low-carrier system CeSb under high pressure studied by simultaneous measurement of electrical resistivity and lattice parameter. \emph{J. Phys. Soc. Jpn.} \textbf{76}, 054706 (2007).

\bibitem{CeBi} 
Rossat-Mignod, J., Burlet, P., Quezel, S., Effantin, J. M., Delac\^{o}te, D., Bartholin, H., Vogt, O. \& Ravot, D. Magnetic properties of cerium monopnictides. \emph{J. Magn. Magn. Mater.} \textbf{31-34}, 398-404 (1985).

\bibitem{LaCeSb} Nakanishi, Y., Sakon, T., Takahashi, F., Motokawa, M., Uesawa, A., Kubota, M. \& Suzuki, T. Fermi surface and magnetic properties of Ce$_x$La$_{1−x}$Sb alloys ($x$=0.5,0.9). \emph{Phys. Rev. B} \textbf{64}, 224402 (2001).

\bibitem{R1} Suzuki, T., Sera, M., Shida, H., Takegahara, K., Takahashi, H., Yanase, A. \& Kasuya, T. In\emph{Valence Fluctuations in Solids} (eds. Falikov, L. M., Hanke, W., \& Maple, M. B.) 255 (North-Holland, Amsterdam, Netherlands, 1981).

\bibitem{R2} Fournier, J. M., \& Gratz, E. Transport properties of rare earth and actinide intermetallics, In \emph{Handb. Phys. Chem. Rare Earths} Vol. \textbf{17} (eds. Choppin, G. R., Eyring, L. \& Lander, G. H.) 409-537 (Elsevier B.V., Amsterdam, Netherlands, 1993).

\bibitem{R3} Suski, W. \& Palewski T. In \emph{Landolt-B\"{o}rnstein - Group III Condensed Matter, Pnictides and Chalcogenides II (Lanthanide Monopnictides)} Vol. \textbf{27B1} (eds. Wijn, H. P. J.) (Springer-Verlag, Berlin, Heidelberg, Germany, 1998).

\bibitem{CanfieldCeSb} Wiener, T. A., \&  Canfield, P. C. Magnetic phase diagram of flux-grown single crystals of CeSb. \emph{J. Alloys Compd.} \textbf{303-304}, 505-508 (2000).


\bibitem{CeSbdHvA} Aoki, H., Crabtree, G. W., Joss, W. \& Hulliger, F. New high frequency dHvA branch of CeSb. \emph{J. Magn. Magn. Mater.} \textbf{97}, 169-170 (1991).

\bibitem{Raquet}  Raquet, B., Viret, M., Sondergard, E., Cespedes, O., \& Mamy, R.  Electron-magnon scattering and magnetic resistivity in 3d ferromagnets.  \emph{Phys. Rev. B} \textbf{66}, 024433 (2002).

\bibitem{Landauer} Landauer, R.  The electrical resistance of binary metallic mixtures. \emph{J. Appl. Phys.} \textbf{23}, 779-784 (1952).

\bibitem{EM_SC} McLachlan, D. S.  Morphology Dependence of the resistivity and Meissner curves in two-phase superconductors.  \emph{Solid State Commun.} \textbf{69}, 925-929 (1989).  

\bibitem{EM_CMR} Kim, K. H., Uehara, M., Hess, C., Sharma, P. A., \& Cheong, S-W. Thermal and electronic transport properties of two-phase mixtures in La$_{5/8-x}$Pr$_{x}$Ca$_{3/8}$MnO$_{3}$.  \emph{Phys. Rev. Lett.} \textbf{84}, 2961-2964 (2000).

\bibitem{MsLachlan} Machlachlan, D. S. An equation for the conductivity of binary mixtures with anisotropic grain structures.  \emph{J. Phys. C} \textbf{20}, 865-877 (1987).  

\bibitem{CeP_MR} Terashima, T., Uji, S., Aoki, H., Qualls, J. S., Brooks, J. S., Haga, Y., Uesawa, A. \& Suzuki, T. Crystal-field $\Gamma_8$-like state in magnetically ordered phases of CeP: its anisotropy and influence on electronic structure via high-field magnetotransport measurements. \emph{J. Phys. Soc. Jpn.} \textbf{70} 3683-3689 (2001).

\bibitem{GdX_MR} Li, D. X., Haga, Y., Shida, H., Suzuki, T. \& Kwon, Y. S. Electrical transport properties of semimetallic GdX single crystals (X=P, As, Sb, Bi). \emph{Phys. Rev. B} \textbf{54} 10483 (1996).

\bibitem{YSb} He, J., Zhang, C., Ghimire, N. H., Liang, T., Jia, C., Jiang, J., Tang, S., Chen, S., He, Y., Mo, S.-K., Hwang, C. C., Hashimoto, M., Lu, D. H., Moritz,  B.,  Devereaux, T. P., Chen, Y. L., Mitchell, J. F., \& Shen, Z.-X.  Distinct electronic structure for the extreme magnetoresistance in YSb. \emph{Phys. Rev. Lett.} \textbf{117}, 267201 (2016).

\bibitem{NdSb} Wakeham, N., Bauer, E. D., Neupane, M. \& Ronning, F. Large magnetoresistance in the antiferromagnetic semimetal NdSb. \emph{Phys. Rev. B} \textbf{93}, 205152 (2016).

\bibitem{Fcalc} Kaneta, Y., Iwata, W., Kasuya, T. \& Sakai. O.  Theoretical calculation for the Fermi surface structures of CeSb in the ferromagnetic and ferrimagnetic AFF1 phases. \emph{J. Phys. Soc. Jpn.} \textbf{69}, 2559-2576 (2000).  

\bibitem{GdPtBi1} Hirschberger, M, Kushwaha, S., Wang, Z., Gibson, Q., Liang, S., Belvin, C. A.,	Bernevig, B. A.,	Cava. R. J. \& Ong, N. P.  The chiral anomaly and thermopower of Weyl fermions in the half-Heusler GdPtBi. \emph{Nat. Mater.} \textbf{15}, 1161-1165 (2016).

\bibitem{GdPtBi2} Suzuki, T.,	Chisnell, R., Devarakonda, A., Liu,	Y.-T., Feng, W., Xiao, D., Lynn, J. W., \& Checkelsky, J. G. Large anomalous Hall effect in a half-Heusler antiferromagnet.  \emph{Nat. Phys.} \textbf{12}, 1119-1123 (2016).

\bibitem{Canfield} Canfield, P. C. \& Fisk, Z. Growth of single crystals from metallic fluxes. \emph{Philos. Mag. B} \textbf{65}, 1117-1123 (1992).

\bibitem{Ames} Materials Preparation Center, Ames Laboratory, US DOE Basic Energy Sciences, Ames, IA, USA, available from: \href{url}{http://www.mpc.ameslab.gov}.


\end{thebibliography}
\end{document}